\documentstyle[twocolumn,aps,eqsecnum]{revtex}

\begin{document}

\thispagestyle{empty}
\setcounter{page}{1}

\title{{\small\begin{flushright} ULB-TH-98/16
\end{flushright}}
Hard thermal loops effective action for 
 $\pi \rightarrow \gamma \gamma$}

\author{Robert D. Pisarski, T. L. Trueman} 
\address{Physics Dept, Brookhaven National Laboratory, Upton, 
NY 11973}

\author{Michel H.G. Tytgat} 

\address{Service de Physique Th\'eorique, CP225 \\
Universit\'e Libre de Bruxelles\\
Bvd du Triomphe, 1050 Bruxelles, Belgium\\
(Talk given
 by M. Tytgat 
at the 5th International
Workshop on \\Thermal Field Theories and their Applications, Regensburg 
(Germany), August 1998)}

\maketitle

\begin{flushright}
  \it To Klaus.
\end{flushright}
\begin{abstract}
$\!\!$
I consider
the low temperature correction to the anomalous coupling 
 of
a neutral pion to two photons from an effective
Lagrangian point of view.
\end{abstract}

\baselineskip=15pt
\pagestyle{plain}

\vskip 1cm

\section{Introduction}

In vacuum, the coupling of a (massless,  on-shell) 
 neutral pion
to two (on mass-shell) photons  is directly 
related to (the coefficient of) the anomalous
divergence of the axial current~\cite{anomaly}.
It is a  truly wonderful thing that the 
axial anomaly (as computed to one-loop in vacuum) is
not renormalized  by higher order quantum
 corrections~\cite{bardeen}. As the anomaly 
emerges from the
 ultraviolet behaviour of quantum fields, 
 this is also the case    
for
thermal corrections~\cite{anot}.

On the other hand, the relation 
between the $\pi^0$ decay
 amplitude and the anomaly can be --and actually is--
subject to thermal corrections~\cite{us}. 
A rather straightforward 
way to verify this is to look at  the anomalous Ward 
identities. 
In these proceedings, however, I
prefer to  follow an effective Lagrangian approach, partly
because it is more suitable for generalisations. 
(See C. Manuel~\cite{cristinaone}, these proceedings.) 

I  use  chiral perturbation 
theory ($\chi$PT)~\cite{cpt}, and thus consider temperatures
(well) below the chiral symmetry breaking scale, 
 $T \ll \sim f_\pi$, with  $f_\pi \approx 97 MeV$ 
the pion
decay constant. Except when explicitely
stated, I 
work in the chiral limit, $m_\pi =0$.
I consider first the pion one-loop quantum corrections, then 
 discuss the low temperature corrections
 and the emergence of 
hard thermal loops. 
This is  only a brief account of the results of~\cite{us}.

\section{VACUUM}
\label{vacuum}

In the vacuum, 
the coupling of a neutral pion to two photons is
described by  the ${\cal O}(P^4)$ effective 
Lagrangian,
\begin{equation}
  \label{eq:vac}
 {\cal L}^{(4)}_{\pi^0 \gamma \gamma} 
=  \left( {e^2 N_c\over 48 \pi^2} \right) 
 \; {1 \over f_\pi}\;
\pi^0 F_{\alpha \beta} \widetilde{F}^{\alpha \beta}
\end{equation}
From~(\ref{eq:vac}), the pion decay amplitude is given by
\begin{equation}
  \label{eq:amp}
  {\cal M} = g_{\pi \gamma\gamma} 
\,\varepsilon_{\alpha\beta\gamma\delta}\,\epsilon^\alpha
\,\epsilon^\beta\, P_1^\gamma P_2^\delta
\end{equation}
with
\begin{equation}
  \label{eq:gpgg}
  f_\pi\, g_{\pi \gamma\gamma} = {e^2 N_c\over 12 \pi^2}.
\end{equation}
and where $\epsilon_{1,2}$ and $P_{1,2}$ are, 
respectively, the
 polarisation vectors and 
 momenta of the outgoing photons.\footnote{
From the effective Lagragian point of view,~(\ref{eq:vac})
is only one of the possible anomalous couplings that arise in 
the gauging of the Wess-Zumino-Witten 
effective action~\cite{wzw}. Most of what I say 
applies straightforwardly to the other anomalous couplings. }
The anomalous 
coupling $g_{\pi\gamma\gamma}$ is independent of the pion 
mass, and  
expressed in term of 
the decay constant $f_\pi$ 
and the RHS of~(\ref{eq:gpgg}), that arises solely from the
anomaly. Without the anomaly and  in the chiral limit, $g_{\pi\gamma\gamma}$ would vanish
to ${\cal O}(P^4)$, which  is consistent with the
 Sutherland-Veltman 
theorem~\cite{sutherland}.

Remarkably, in the chiral
limit, $m_\pi = 0$, and for photons on  mass-shell,
$P_1^2 = P_2^2 =0$, the relation~({\ref{eq:gpgg})
is   exact~\cite{anomaly}.
This result rests 
on the non-renormalisation  of the 
axial anomaly~\cite{bardeen}, and can be derived using the
anomalous Ward identities only~\cite{jackiw}.
It also has a translation  
in  the effective Lagrangian language~\cite{donoghue}.
Note first that the effective
coupling~(\ref{eq:vac}) 
saturates the anomaly, {\em i.e.}  
correctly reproduced the anomaly of  the underlying quark
degrees of freedom. 
Under $\pi/f_\pi \rightarrow
\pi/f_\pi + \alpha$, 
\begin{equation}
  \label{eq:noether}
  \delta {\cal L}_{\pi^0 \gamma \gamma} = \alpha \;
\left( {e^2 N_c\over 48 \pi^2} \right) 
 \;F_{\alpha \beta} \widetilde{F}^{\alpha \beta}
\end{equation}
so that
\begin{equation}
\label{eq:anodiv}
\partial^\alpha J_{5,\alpha}
= - \frac{e^2 N_c}{48 \pi^2} F_{\alpha \beta} 
\widetilde{F}^{\alpha \beta} 
\end{equation}
with $
 J_{5, \alpha} = f_\pi \partial_\alpha \pi^0 + \ldots$, in term
of the pion field.
This implies in particular that higher order terms in 
the  momentum expansion of $\chi$PT, and 
thus quantum corrections to~(\ref{eq:vac}), are
 non-anomalous.
According to the power counting of $\chi$PT, the 
anomalous Lagrangian~(\ref{eq:vac}) is ${\cal O}(P^4)$, so that
one-loop
 corrections  with  insertions of one  anomalous and
 one ${\cal O}(P^2)$ vertex  induce terms that are
 ${\cal O}(P^6)$. One such term is
 \begin{equation}
   \label{eq:nonano}
   {\cal L}^{(6)}_{\pi^0 \gamma \gamma} \propto  
{N_c \alpha \over f_\pi^2} 
\varepsilon_{\mu\nu\alpha\beta} \partial_\lambda F^{\mu\nu} 
F^{\alpha\lambda} \partial^\beta \pi_0 + \ldots
 \end{equation}
If the photons are on  mass-shell, as is clear 
for dimensional reason, all such operators can only give 
a contribution  that is proportional
to the pion mass. Thus, to 
${\cal O}(P^6)$  
\begin{equation}
  \label{eq:mod}
  f_\pi\, g_{\pi \gamma\gamma} = {e^2 N_c\over 12 \pi^2} + 
{\cal O}(m_\pi^2)
\end{equation}
on mass-shell, wich implies that~(\ref{eq:gpgg}) is exact in the
chiral limit, $m_\pi^2 =0$. \footnote{In other words, the 
${\cal O}(P^4)$ effective Lagrangian~(\ref{eq:vac}) is the only possible term 
that is  compatible with the violation of the Sutherland-Veltman
theorem induced by the anomaly.}

\section{Thermal bath}
\label{lowt}

The discussion of the last section
suggests that non-trivial corrections to the 
pion coupling can arise at finite temperature -- despite the
fact that the anomaly itself is not affected --
for   
the presence of a  thermal bath introduces, beside the 
pion and photon momenta,  a new 
scale set by the temperature, $T$. Temperature can modify 
the relation~(\ref{eq:gpgg}), pretty much like the pion mass 
does, Eq.~(\ref{eq:mod}). Of course, unlike a pion
 mass, finite temperature 
  preserves chiral symmetry. What will eventually matter here
 is that manifest Lorentz invariance is
 lost  in a  thermal bath.

First, the
power counting of $\chi$PT has to be modified to also take into account
the powers of $T$.  
Consider for instance the
 correction
to the pion decay 
constant $f_\pi$, that enters in~(\ref{eq:vac}). 
In the vacuum,
the one-loop correction 
(corresponding to  a tadpole diagram) 
 is a homogeneous function of the pion mass,  
${\cal O}(m^2_\pi)$. Thus, in the chiral limit, 
$f_\pi^{1-loop} = f_\pi^{\rm bare}$. (Actually, this
is true to all orders.) 
At finite temperature however, the  
correction is  ${\cal O}(T^2)$, for $m_\pi \ll T$,  and 
\begin{equation}
\label{eq:fpit}
f_\pi(T) = ( 1 - T^2/12 f_\pi^2)\; f_\pi,
\end{equation}
for two light quark flavours~\cite{leut}.
The new scale set by the temperature induces
non-trivial corrections, even in the chiral
 limit. If one na\"{\i}vely 
substitutes $f_\pi(T)$ in~(\ref{eq:vac}),
one gets from~(\ref{eq:gpgg}) a coupling 
$g_{\pi \gamma\gamma}(T)$ that increases with temperature, 
which is a strange -- but not 
inconceivable-- result. However, this low 
temperature behaviour can be contrasted to 
the one near $T_c$,
obtained using a linear sigma model with 
constituent quarks~\cite{rob}: in the chiral
limit, the anomalous
coupling vanishes above $T_c$.

At finite, but low temperature, the expansion of the
effective Lagrangian is thus in 
powers of $m_\pi^2$, $T$ and
the external momenta. The 
thermal correction to the pion decay constant is 
${\cal O}(T^2)$ and arises from pions of the thermal
bath with momentum $P \sim T$. This is similar to the
thermal corrections --known as hard thermal 
loops (HTL)-- that arise in hot gauge theories~\cite{BP}.
This suggests to consider the following hierarchy of
scale:
\begin{equation}
  \label{eq:hier}
  P \ll T \ll \sim f_\pi,
\end{equation}
where $P$ stands for  external momenta. 
Then, as the leading contribution to the anomalous coupling
is ${\cal O}(P^4)$, the low temperature  
correction to the pion decay constant
 in~(\ref{eq:vac}) gives a contribution that is 
 ${\cal O}(P^4 T^2)$, and {\em a priori} dominates over the  ${\cal O}(P^6)$ 
 quantum 
corrections (and {\em a fortiori}  the  ${\cal O}(P^4 m_\pi^2)$ ones that
vanish in the chiral limit). This leads us
to consider the  ${\cal O}(T^2)$
 effective action of soft pions, $P \ll T, f_\pi$,   
in a cool thermal bath, $T\ll \sim f_\pi$. (As 
the number density of these soft
pions in the thermal bath is $n \sim T/p \gg 1$, this seems like
a reasonable thing to consider.) To ${\cal O}(P^4 T^2)$
there are {\em a priori} many more 
 terms that one can write
that do not vanish in the chiral limit. Not only is 
manifest Lorentz invariance explicitely
broken --which gives  
more  {\em local} terms-- but the non-analyticities,
characteristic of finite $T$  amplitudes,
transcribe into  {\em non-local} effective couplings, as is
well-known from  hot gauge theories~\cite{robandi}.  
Calculations to one-loop in the HTL 
approximation~\cite{us} reveal that 
the complete, ${\cal O}(T^2)$,  effective action is
\begin{equation}
  \label{eq:htl}
 {\cal L}_{\pi^0 \gamma \gamma}(T) 
=  \left( {e^2 N_c\over 48 \pi^2} \right) 
 \; {1 \over f_\pi(T)}\;
\pi^0 F_{\alpha \beta} \widetilde{F}^{\alpha \beta}
\end{equation}
$$
 -\; \frac{T^2}{12 f_\pi^2} \left( {e^2 N_c\over 48 \pi^2} \right)
\int \frac{d \Omega_{ \hat{k} } }{4 \pi} \; 
H_{\gamma \alpha}
\frac{ \hat{K}^\alpha \hat{K}^\beta}{- (\partial \cdot \hat{K})^2}
F_{\gamma \beta} \; ,
$$
where 
$\widetilde{F}^{\alpha \beta} =
\epsilon^{\alpha \beta \gamma \delta} F_{\gamma \delta}/2$, 
$
H_{\alpha\beta} = \partial_\alpha H_\beta - \partial_\alpha H_\beta 
$,
and
\begin{equation}
H_\alpha = {1\over f_\pi}\varepsilon_{\alpha\beta\gamma\delta} 
F_{\beta\gamma} \partial_\delta \pi^0 \; .
\label{dk}
\end{equation}
The vector $\hat K = ( i, \hat{k})$; one  integrates
over all angles $\hat{k}$.  This integration effectively 
represents
the hard, massless pions in the one loop integral.
One  can verify that the second term of~(\ref{eq:htl}) is
indeed
invariant under 
an axial transformation,
 $\pi/f_\pi(T) \rightarrow \pi/f_\pi(T)  + \alpha$, and thus,
as required,  is non-anomalous.\footnote{That the thermal corrections to the
  WZW effective action are non-anomalous has been proven in~\cite{estrada}.} 
It nevertheless contributes to the effective anomalous coupling.
Together with the correction to the pion decay constant in the 
 first term of~(\ref{eq:htl}), it finally
gives
\begin{equation}
  \label{eq:couplt}
  g_{\pi\gamma\gamma}(T) = \;(1 - T^2/12 f_\pi^2)  \;
g_{\pi\gamma\gamma}
\end{equation}
The coupling thus decreases with temperature, a result
that -- if nothing else --
 is consistent with the behaviour found 
near $T_c$~\cite{rob}.
It is quite interesting that the coupling decreases with $T$
precisely 
like $f_\pi(T)$. I don't know whether this is a coincidence.

\section{CONCLUSIONS}

Further understanding   of the results sketched in the
previous sections  can 
be gained from an analysis of the anomalous Ward identities
at finite temperature~\cite{us}. 
For instance, explicit one-loop  calculations allow to 
see that, unlike in vacuum, in a thermal bath 
the anomalous divergence of the axial current is 
not saturated by the one-pion pole. (This is what allows
the temperature dependence of the effective coupling to be 
non-trivial.) Also, the decay of a neutral pion into two photons is
 only one particular
anomalous amplitude. The complete HTL contribution  
to the
Wess-Zumino-Witten (gauged) effective action have been 
computed by C. Manuel~\cite{cristina}. (For further application of 
HTL in chiral dynamics, see also~\cite{manuelbis}.)

I  would like to emphasise that the rather 
academic problem discussed here 
-- the effective anomalous coupling of a neutral pion
at finite temperature  --
is  only one simple
aspect of  the broader class of problems  dealing with  
the relation between  axial anomalies and their
phenomenological manifestation  at  finite temperature.
For example,  shifts in anomalous couplings
of hadronic states, like the $\omega$ and $\phi$, could be relevant
in heavy ion collisions. 
Another instance  is offered by  the 't Hooft anomaly matching
conditions~\cite{thooft}, for which  manifest  Lorentz
covariance is a crucial ingredient~\cite{coleman}; finite temperature could
allow for more exotic solutions, maybe parity 
doublets~\cite{detar}.
A very interesting problem concerns the relation, at finite temperature,
 between
the breaking of  the $U(1)_A$
symmetry and the meson spectrum, in particular
 the  $\eta^\prime$. 
For three or more flavours, effective
Lagrangian~\cite{frank}
and instantons~\cite{shushu} arguments suggests
 and effective~\footnote{Effective means here that
  violation of $U(1)_A$ is relegated to irrelevant operators.}
restoration of the $U(1)_A$ symmetry at the critical
temperature of chiral symmetry breaking. Two flavours is 
marginal
but a large $N_c$ argument seems to lead to similar 
conclusions~\cite{dima}. 


\begin{references}
%
\bibitem{anomaly}
S.L. Adler, Phys.Rev. {\bf 177}, 2426 (1969);
J.S. Bell, R. Jackiw, Nuovo Cim. {\bf 60A}, 47 (1969).
%
\bibitem{bardeen}
S. L. Adler and W.A. Bardeen, Phys. Rev. {\bf 182} (1969), 1517.
%
\bibitem{anot}
L. Dolan and R. Jackiw, Phys. Rev. {\bf D9} (1974), 3320;
M. Reuter and W. Dittrich, Phys. Rev. {\bf D32} (1985), 513;
F. Ruiz Ruiz and R. 
Alvarez-Estrada, Phys. Lett. {\bf B180}  (1986), 153;
A. Das and A. Karev, Phys. Rev. {\bf D36}  (1987), 623;
R. Baier and E. Pilon, Zeit. fur Phys. {\bf C52} (1991), 339;
H.~Itoyama and A.~H. Mueller, Nucl. Phys. {\bf B218}  (1983), 
349.
%
\bibitem{us}
R.D. Pisarski, T.L. Trueman and M.H.G. Tytgat, Phys. Rev. 
{\bf D56}, 7077 (1997); R. D. Pisarski, 
T.L. Trueman and M.H.G. Tytgat, hep-ph/9804466.
%
\bibitem{cristinaone}
C. Manuel, hep-ph/9809273.
%
%
\bibitem{cpt}
S. Weinberg, {\it Physica} {\bf 96 A} (1979) 327;
 J. Gasser and H. Leutwyler, {\it Nucl. Phys.} {\bf B250}
(1985) 465; 
J.~F.~Donoghue, E.~Golowich, and B.~R.~Holstein, ``{\it Dynamics
of the Standard Model}", Cambridge 
University Press,  New York 1996.
%
\bibitem{jack}
J. Steinberger, 
 Phys.Rev. {\bf 76}, 1180 (1949).
%

\bibitem{wzw}
J. Wess and B. Zumino, Phys. Lett. {\bf B37}, 95 (1971);
E.~Witten, {\bf B223}, 422 (1983).
%
%
\bibitem{jackiw}
See, for instance, R.~Jackiw, in
{\it Current Algebra and Anomalies}, 
(Princeton Univ. Press, Princeton, 1985).
%
\bibitem{sutherland}
D. G. Sutherland, Nucl. Phys. {\bf B2}, 433 (1967); 
M. Veltman, Proc. R. Soc. London {\bf A301}, 107 (1967).
%
\bibitem{donoghue}
J. F.~Donoghue and D. Wyler, Nuc. Phys. {\bf B316},289 (1989).
%
\bibitem{leut}
P. Binetruy and M. K. Gaillard, 
Phys. Rev. {\bf D 32}, 931 (1985);
J. Gasser and H. Leutwyler, Phys. Lett. {\bf B 184}, 
83 (1987);
M. Dey, V. L. Eletsky, and B. L. Ioffe, ibid. 252, 620 
(1990);
A. Bochkarev and J. Kapusta, Phys. Rev. {\bf D 54},
 4066 (1996); 
R. D. Pisarski and M. Tytgat, Phys. Rev. {\bf D 54},
 2989 (1996). 
%
%
% 
\bibitem{rob}
R.D. Pisarski, Phys. Rev. Lett. {\bf 76}, 3084 (1996);
R. Baier, M. Dirks, and O. Kober, Phys. Rev. {\bf D54}, 
2222 (1996); F. Gelis, hep-ph/9806425;
L.L. Salcedo, hep-th/9807221. 
%
\bibitem{BP}
R.~D.~Pisarski, {\rm Phys. Rev. Lett.} {\bf 63}, 1129 (1989);
E.~Braaten and 
R.~Pisarski, {\rm Nucl. Phys.}  {\bf B337}, 569 (1990);
J.~Frenkel 
and J.~C.~Taylor, {\rm Nucl. Phys.}  {\bf B334}, 199 (1990).
%
\bibitem{robandi}
R.D. Pisarski and M. Tytgat, Phys. Rev. Lett. {\bf 78}, 
3622 (1997).
%
\bibitem{estrada}
R.F. Alvarez-Estrada, A. Dobado, A. Gomez Nicola, Phys.Lett. {\bf B319}, 238 (1993).
\bibitem{cristina}
C. Manuel, Phys.Rev. {\bf D57}, 2871 (1998). 
%
\bibitem{manuelbis}
C. Manuel, Phys.Rev. {\bf D58}, 016001 (?)  (1998); C. Manuel and 
N. Rius, hep-ph/9806385.
%
\bibitem{thooft}G. 't Hooft, in {\it Recent 
developments in gauge theories}, ed. G. 
't Hooft {\em et al}. (Plenum Press, N.Y., 1980).
%
\bibitem{coleman} 
S. Coleman and B. Grossman,  Nucl.Phys. {\bf B203}, 205 
(1982);
T. Banks, S. Yankielowicz,  A. Schwimmer, Phys.Lett. 
{\bf 96B}, 67 (1980). 
%
\bibitem{detar}
C. DeTar, T. Kunihiro,  Phys.Rev. {\bf D39}, 2805 (1989). 
%
\bibitem{frank}
R. D. Pisarski, F. Wilczek, Phys.Rev. {\bf D29}, 338 (1984). 
%
\bibitem{shushu}
T. Schafer, E.V. Shuryak,  Rev.Mod.Phys. {\bf 70}, 
323 (1998); N. Evans, S. D.H. Hsu, M. Schwetz, Phys. Lett. {\bf B375}, 262 (1996). 
%
\bibitem{dima}
D. Kharzeev, R.D. Pisarski, M. H.G. Tytgat,  Phys.Rev.Lett.
{\bf 81}, 512 (1998)

\end{references}
\end{document}